\documentstyle[12pt]{article}

\begin{document}

\title{ON THE CRITICAL BEHAVIOUR OF HEAT CONDUCTING SPHERE 
OUT OF HYDROSTATIC EQUILIBRIUM.}
\author{L. Herrera\thanks{
On leave from Departamento de F\'\i sica, Facultad de Ciencias,
Universidad Central de Venezuela, Caracas, Venezuela and Centro de
Astrof\'\i sica Te\'orica (C.A.T.), M\'erida, Venezuela.}\\
\'Area de F\'\i sica Te\'orica, Facultad de Ciencias, \\
Universidad de Salamanca, \\ 37008 Salamanca, Espa\~na.\\
and
\and
J. Mart\'\i nez\\
Grupo de F\'{\i}sica Estad\'{\i}stica, Departamento de F\'\i sica, \\
Universidad Aut\'onoma de Barcelona,\\ 08193 Bellaterra, Barcelona, Espa\~na.}
\date{}
\maketitle

\begin{abstract}
We comment further on the behaviour of a heat conducting fluid when a 
characteristic parameter of the system approaches a critical value
\end{abstract}

\newpage

In a recent paper \cite{DiHe97} (hereafter referred to as I) we have obtained 
that for a heat conducting sphere, immediately after leaving the hydrostatic
equilibrium, the following equation is satisfied
\begin{equation}
 -e^{(\nu-\lambda)/2}R=(\rho+P_r)(1-\alpha)\times\dot{\omega},
\label{pseudonewton}
\end{equation}
where a dot stands for partial differentiation with respect to time,
$\omega$ is the velocity of matter as measured by a Minkowskian
observer, and $\alpha$ 
is defined by 
\begin{equation}
\alpha=\frac{\kappa T}{\tau(\rho+P_r)},
\label{alfa}
\end{equation}
being $\tau$ the relaxation time for thermal signals, and $\kappa$
the thermal conductivity coefficient.

Equation (\ref{pseudonewton}) was obtained from a linear
perturbative scheme where $\dot{\omega}$ and $R$ (as well as
$\omega$, heat flow, and time derivatives of physical variables) are
small quantities of first order.  
Interpreting eq.(\ref{pseudonewton}) as "Newtonian" equation
$$
\mbox{Force=mass}\times\mbox{acceleration,}
$$
it appears that the "effective inertial mass" decreases as $\alpha$
increases from zero, and vanishes at the critical point defined by
condition $\alpha=1$.
On the other hand, from the analysis of stability and causality in
dissipative relativistic fluids \cite{HiLi83}, it follows that
causality and hyperbolicity (which imply stability) require for
dissipative viscous free systems
\begin{equation}
\label{tau1}\tau >\frac{\kappa T}{\rho +p}+\left( \frac \kappa n\right) 
\frac{c_s^2}{c_p}
\end{equation}
\begin{equation}
\label{tau2}\tau >\left( \frac{\kappa T}{1-c_s^2}\right) \left[ \frac 1{\rho
+p}+\left( \frac 1{nT}\right) \left[ \frac 1{c_v}-\frac{c_s^2}{c_p}\right] -%
\frac{2\alpha_p }{nc_v\kappa _T(\rho +p)}\right] 
\end{equation}
and 
\begin{equation}
\label{tau3}\tau >\frac \kappa {nc_s^2c_v}\left[ \frac{2\alpha_p T}{\kappa
_T(\rho +p)}-1\right] 
\end{equation}
where $n, c_s, c_p, c_v, \kappa_T,$ and $\alpha_p$ denote the
particle density, the sound speed, the specific
heat at constant pressure and volume, the thermal expansion
coefficient and the isothermal compressibility respectively.
These expressions are found from eq.(146)-(148) in \cite{HiLi83},
taking the limit $\beta_o,\beta_2\rightarrow\infty$ and
$\alpha_i=0$ (this method was applied in \cite{Maartens96} to the
case in which only bulk viscous preturbations were present).
It should be kept in mind that conditions above, are obtained within a linear
perturbative scheme.

Obviously, condition (\ref{tau1}) is violated at the critical point
(in fact it is violated, slightly below it, see comment at the end).
However it is not difficult to find physical
conditions for which the numerical values of variables entering in
the definition of $\alpha$ lead to $\alpha=1$. Therefore, the
relevant question is: Can a physical system actually reach the
critical point? . If the answer to this question is negative, then it
should be explained how a given system avoids the critical point.
Since, as mentioned before, numerical values of $\kappa$, $T$,
$\tau$, $\rho$, and $P_r$; leading to $\alpha\approx 1$ may
correspond to a non very exotic scenario. 
On the other hand, a positive answer seems to be prohibited by
causality and stability conditions. However, we shall conjecture here
that this might not be the case. In fact, what eq.
(\ref{pseudonewton}) tell us is, that as we approach the critical
value, a linear perturbation scheme fails, since for
$\alpha\approx1$, $\dot{\omega}$ will not longer be a small quantity
of first order, invalidating thereby the linear approximation used to
obtain (\ref{pseudonewton}-\ref{tau3}).

In other words, vanishing of the "effective inertial mass" at the
critical point indicates that linear approximation is not valid at
that point. So it seems that the behaviour of the system close to the
critical point can't be studied with a linear perturbative scheme,
but requires the integration of the full system of dynamic equations.
Thus, our conjecture is that far below the critical point, where the
perturbative scheme is valid, the "effective inertial mass" decreases
as $\alpha$ increases. Close to the critical point, the linear
approximation ceases to be valid, and the "effective inertial mass"
is not longer given by $(\rho+P_r)(1-\alpha)$. An exact numerical
example of a system passing through the critical point during its
evolution will prove (or disprove) our conjecture. 

It is worth noticing that situations where the critical point is
reached for very small relaxation time (much smaller than any
relevant timescale of the system) are deprived of physical interest,
since eq.(\ref{pseudonewton}) is valid on a timescale of the order of
relaxation time.

Finally, observe that, condition (\ref{tau1}) is violated
below, but very close to, the critical point for small values of $c_s^2$.

\section*{Acknowledgments}

This work was partially supported by the Spanish Ministry of Education under
Grant No. PB94-0718

\end{document}